# Kurt Lewin, psychological constructs and sources of brain cognitive activity.

**Włodzisław Duch**

Department of Informatics, Faculty of Physics, Astronomy and Informatics,
and Neurocognitive Laboratory, Center for Interdisciplinary Modern Technologies,
Nicolaus Copernicus University, Toruń, Poland.

**Abstract.** Understanding mind-brain-environment relations is one of the key topics in psychology. Kurt Lewin, inspired by theoretical physics, tried to establish topological and vector psychology analyzing patterns of interaction between the individual and her/his environment. The time is ripe to reformulate his ambitious goals, searching for ways to interpret objectively measured brain processes in terms of suitable psychological constructs. Connecting cognitive and social psychology constructs to neurophenomics, as it is done now in psychiatry, should ground them in physical reality.

**Keywords**: psychological constructs, dynamical systems, neurodynamics, neuropsychiatric phenomics, attractor networks, Research Domain Criteria, Kurt Lewin.

## 1. Searching for bridges

Psychology is based on constructs that were derived from common-sense understanding of mental processes and behavior, refined over the years, but without connection to physical processes in the brain. Paul Churchland in 1981 article "Eliminative Materialism and the Propositional Attitudes" argued that common-sense understanding of the mind (or folk psychology) is false and that progress in understanding brain functions will lead to elimination of most concepts that psychology is based on. Patricia Churchland in an influential book "Neurophilosophy" (1986) has also supported eliminativist position. After all biology, chemistry and physics had to abandon many concepts and theories that in the past seemed reasonable, but were eliminated by deeper understanding of biological and physical processes. Many constructs introduced by Freudian psychology have been rejected. Other constructs, such as distinction between different types of memory were gradually introduced: working memory (~1960), episodic and semantic memory (1972), implicit memory (1981). Such constructs may in future still be splitted into more subtypes that correspond to specific brain processes.

Defining psychological constructs is harder than physics or biology constructs. Individual variability of brains, irreversible influence of experiments on cognitive systems, makes comparison and stability of results quite difficult. For that reason Smedslund (2016) concluded that psychology cannot be an empirical science. Failure of structuralism and move towards behaviorism created an interest in conceptual foundations of psychology, following the success of physics. Kurt Lewin has published several influential books (1936, 1938, 1951) introducing psychological force field analysis, describing psychological processes in topological spaces, focusing on conceptual representations and measurements of psychological forces. Daniel Khaneman in his Nobel Prize speech (2002) said: "As a first-year student, I encountered the writings of the social psychologist Kurt Lewin and was deeply influenced by his maps of the life space, in which motivation was represented as a force field acting on the individual from the outside, pushing and pulling in various directions. Fifty years later, I still draw on Lewin's analysis of how to induce changes in behavior …".

Lewin has formulated field theory to describe behavior as a result of patterns of interactions between the individual and the environment. In his theory cognitive dynamics is represented as a movement in phenomenological (he has used the word “hodological”) space, a “life space” or a field that includes person's values, needs, goals, motives, moods, hopes, anxieties, and ideals. Forces in this field arise in social situations, driving cognitive movement toward or away from goals of the person. His description of the process of mental change include 3 stages: unfreezing or escaping the inertia, transition without clear idea where it leads, and freezing or crystallizing new behaviors. These ideas can be linked to activity of neural networks, simulated in computer or observed in neuroimaging (Duch, 2010;2011).

Perhaps the time is ripe to take ideas of Kurt Lewin seriously. I will first make a short review of different trends in psychology that were initiated by Lewin’s work, and then present some developments in neuroscience that should ground psychological constructs in objectively measured brain processes.

## 2. Dynamical ideas in psychology

Lewin’s “field theory” ideas, discussed in the next section, led to the development of Gestalt approach, avoidance conflict model, and personality psychology (Trempała, Pepitone, Raven 2006). They have contributed to many trends in science that are briefly presented below.

George Kelly (1955) has formulated personal construct psychology, as complete theory of cognition, action, learning and intention, using geometry of psychological spaces as alternative to logic. This may be considered as a simplification of Lewin’s field theory, making it easier to apply in practice. Subjective reality is expressed in terms of psychological constructs. Kelly (1955) assumed that "A person's construction system is composed of a finite number of dichotomous constructs." These constructs provide psychological dimensions that characterize people and mental events. Understanding people requires identification of constructs they use, defined by results of psychometric tests, mental states, and various other types of elements. This information is stored in a “Repertory Grid” matrix, with rows representing personal constructs that are relevant to some specific purpose, and columns that represent various elements, such as opinions, preferences, or potential actions. Techniques based on personal construct psychology (PCP) have wide applications in social psychology, psychotherapy, personality assessment and human resources in business context, supporting decision-making processes and helping to study personal and interpersonal systems of meaning (Abraham, Shaw, 1984). WebGrid (Gains and Shaw, 1997; http://webgrid.uvic.ca) approach has further developed the conceptual representation system, using graded constructs instead of simple dichotomies, providing interactive software tools to elicit and analyze mental models of individuals and groups within specific domains of experience.

Many other grid-based software approaches were created to represent psychological constructs. Perceptions, beliefs, values, personality traits and other constructs may be represented in high-dimensional spaces, with measured or estimated values in particular dimensions. Roger Shepard’s research on mental representations (Shepard 1987; 1994) was focused on understanding topology of such spaces, searching for interesting regular structures and invariants that represent universal psychological laws. Using non-metric multidimensional scaling approach he has shown how physical properties of stimuli, important from evolutionary point of view, are reflected in mental models. Perception of objects, spatial relations, color space, color constancy, the pitch of sounds, tastes, and abstract numbers rely on neural transformations that support optimal generalization and categorization. Many experiments with human and animals perception proved that the use of inter-stimulus distances $D$ to measure similarity of stimuli leads to exponentially decaying generalization probabilities $P(D)=\exp(-\alpha D)$ of behav-ioral reactions. Another universal psychological law says that the discriminative reaction time falls off as the inverse of the inter-stimulus distance (Shepard 1987; 1994) Both laws can be derived from optimal Bayesian principles. Psychological laws related to perception and mental representation evolved over a long time, reflecting universal physical principles. For example, 24-hour circadian rhythm is a consequence of the law of angular momentum conservation that warrants stable spinning of the Earth. Similarity and causal relations between mental representations of physical events are isomorphic to the relations between physical events themselves, although there is obviously no direct resemblance of mental representations to physical events. This second-order similarity is the basis for efficacy of learning by mental simulations of physical events, thought experiments, observed in mental rotation and imagery. Apparent motion is constrained by kinematic geometry in three-dimensional space, and reflects anticipations about object constancy.

Geometric/dynamical ideas related to mental models may be found in many fields. Lewin's ideas inspired decision field theory (Busemeyer & Townsend, 1993) and Discrete Process Model (DPM) theory (Rainio, 2009). In DPM psychic forces are defined as the probability of transition from one cognitive state in valence field to another. Rainio concluded "It seems obvious that Kurt Lewin's brilliant intuitive insights concealed fundamental ideas which lead to new understanding not only in dynamical psychology but also in a much greater domain of philosophy".

Cognitive science book "Mind as motion" (Port and van Gelder, 1995) focused on dynamical systems approach as a general framework for theories of cognition, discussing developmental processes, language, articulation in speech, decision making, perception, learning, spatial orientation and many other issues. Computational systems used by symbolic cognitive science models belong to a restricted subclass of dynamical systems. Most psychological processes change in a continuous way and cannot be represented by computational systems. Conceptual representation using discrete symbols, verbal descriptions of nonlinear dynamics, may not be a good approximation of behavior. Early cognitive science relied on symbolic artificial intelligence in search for unified theories of cognition (Newell, 1994). Reasoning and problem solving were understood as search in discrete problems spaces. Several cognitive architectures (ACT-R, SOAR, CLARION and others) were proposed as computational models explaining many aspects of cognition (see the review in Duch, Oentaryo and Pasquier 2008). This approach had some successes, but it was never useful in understanding perception, motor control or imagery. Recent revival of artificial intelligence based on machine learning, deep neural networks and dynamical systems has led to great progress in technical applications. These approaches should be fully encapsulated in new cognitive architectures that will advance connectionist models to a new level, providing simulations of many behavioral functions. Generic processes of self-organization and learning (Kelso 1997) lead to creation and evolution of complex patterned behavior that can be analyzed in psychological spaces.

Dynamical approach has also been used in developmental psychology to describe grasping, crawling and learning to walk (Thelen and Smith,1996; Smith and Thelen 1993). It can explain many aspects of language, including some features of semantics and conceptual integration (Fauconnier, 1994), stream of thoughts represented by trajectories in psychological spaces (Elman, 1995), and development of natural categories and associations in language in spaces created by latent semantic analysis (Landauer and Dumais, 1997). In "The Continuity of Mind" (Spivey, 2007) a step towards connecting psychology with brain activity is made, by considering trajectories through the neural state space. This approach was applied to motor action, vision, formation of categories, language, memory and problem solving.

Many recent discoveries in neuroscience show that interpreting brain neurodynamics in language of dynamical systems leads to a deeper understanding of current psychological con-structs, and to creation of the new ones, more specific and closely linked to the physical reality of objectively measured brain activity. Symbolic, conceptual description of continuous processes may in most cases be useful as a rough description of behavior. However, more detailed dynamical models of mental processes will be hard to describe conceptually. Some aspects of nonlinear dynamics may be visualized in psychological spaces.

I have formulated definition of mental forces (Duch, 1996; 1997; 2012) similar to the Discrete Process Model of Rainio (2009), pointing out that such forces should be measured by the probability of transitions between brain states in neurodynamics. We can measure brain activity using many techniques, such as EEG, MEG, NIRS, PET, fMRI and other approaches. How this neural activity is spread through the connectome to various regions of the brain, and how joint activity of these regions is related to behavior is not yet clear. Even worse, description of the phenomenology of mental states through introspections seems impossible, as shown for example by Hurlburt and Schwitzgabel (2007), and Schwitzgabel (2011). We can describe only those mental states that correspond to strong, repeatable brain activations associated with linguistic tokens (Duch, 2012). Neurophenomenology proposed by Varela (1996) explores mutual constraints between brain activity and inner experience. Still most of neuroscience and neuroimaging research ignores subjective experience, while psychological theories forget about the brain processes behind theoretical constructs they postulate.

## 3. Kurt Lewin's field theory and attractor neural networks

Lewin was especially concerned with "A dynamic theory of personality" (1935), including all factors that may influence behavior. His famous equation B=f(P,E) expresses behavior B as a function of both the person P and the environment E. More precisely (Lewin 1951), behavior of a person depends on the: genetic and other factors that contribute to the brain structure on which personality of this person develops in a given environment; dynamic approach that involves forces determining actions; psychological perspective of a person subjectively perceiving her/his "life space", relevant internal-external factors; analysis of the situation, reflection, associations, understanding; finally behavior as a function of the total field containing all these elements changing in time, described in topological spaces divided into some regions. This has been illustrated in his book using a diagram that divides the field into regions and arrows representing forces based on valence. Fig. 1 represents positive central force field G (Lewin 1938, Fig. 33) and different regions that have influence on this central region. A person P placed in region A exerts some force on C, denoted as $f_{A,C}$. This is still a metaphoric description of intentional activity and goal-directed behavior.

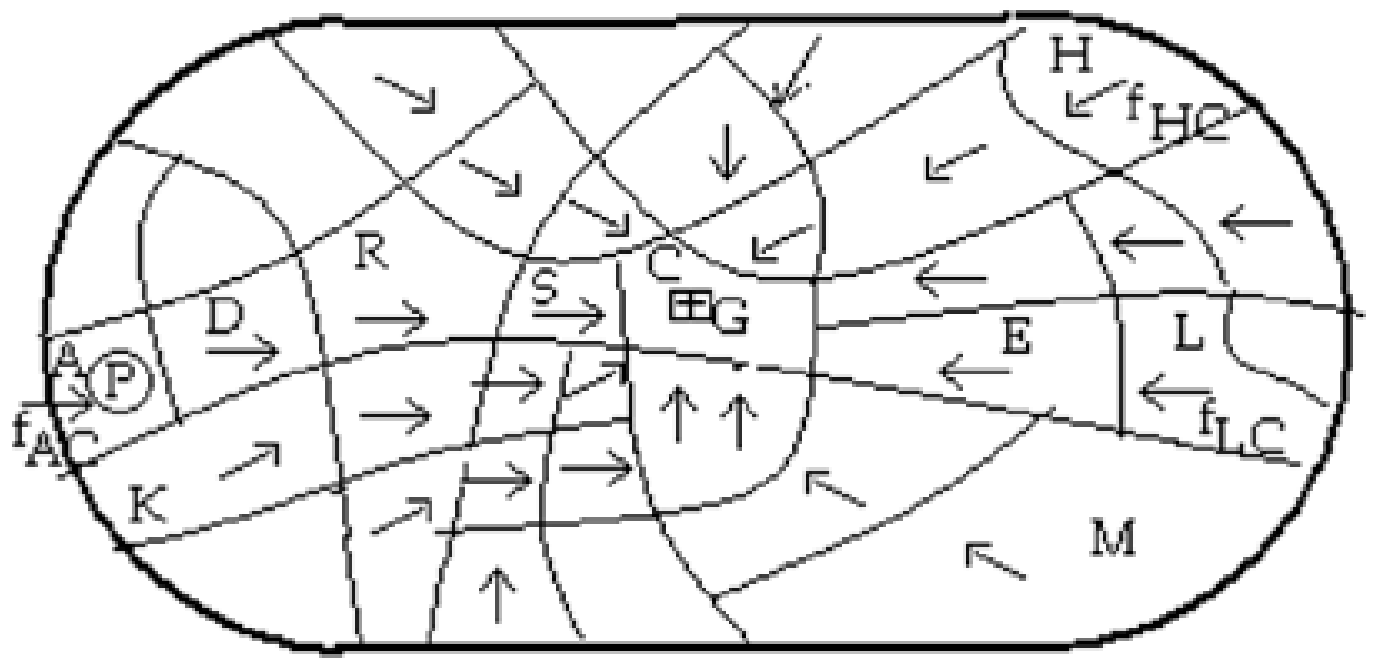

Fig. 1. Illustration of Lewin's force field.

According to Lewin psychology should use constructs to represent causal influences and connections between observations. He has introduced several new constructs, such as valence, action research, sensitivity training, group dynamics, mind as a complex energy field, behavior as a change in the state of this field, regions, life space, forces and tension, equilibrium states. Complex energy field can be presented in the language of dynamical systems. Transformation between states of activations in neural space S(N) and between mental states S(M) described in psychological spaces is already to some degree possible. Brain-computer interfaces (BCI) analyze and interpret mental activity, changing it into intentional actions. Mind reading is an exciting and rapidly developing field. Mapping from Brain ⇔ Mind, or Objective ⇔ Subjective, may be represented in symbolic form as:

$$S(B;E) \Leftrightarrow S(M;E'),$$

where environment E is reduced at the mental level to E', psychological perspective of E, as Lewin has noticed. "Genetic factors" forming the foundations on which personality may develop are now greatly expanded.

National Institute of Mental Health (NIMH) in the USA has initiated an ambitious Research Domain Criteria (RoDC) approach to multi-level neuropsychiatric phenomics (Insel et al. 2010). Instead of traditional description of mental disease by listing their symptoms deregulation of normal activity of 5 large brain systems is considered: negative/positive valence systems, arousal-regulatory system, cognitive system, and social processes system. These systems are characterized at many levels by genes, molecules, cells, circuits, physiology at the physical level, and description of behavior, plus subjective self-reports, collected using various research paradigms. While all physical levels influence behavior it is the activity at the neural system level, representing Lewin's complex energy field, that is directly responsible for action. Neurodynamics explains cognition and behavior, it is measured using neuroimaging techniques, and can also be simulated computationally using neural network models. There is no simple causality here, as environment and behavior may influence genetic level, changing the system through epigenetic regulation of gene expressions. The causal chain includes a loop:

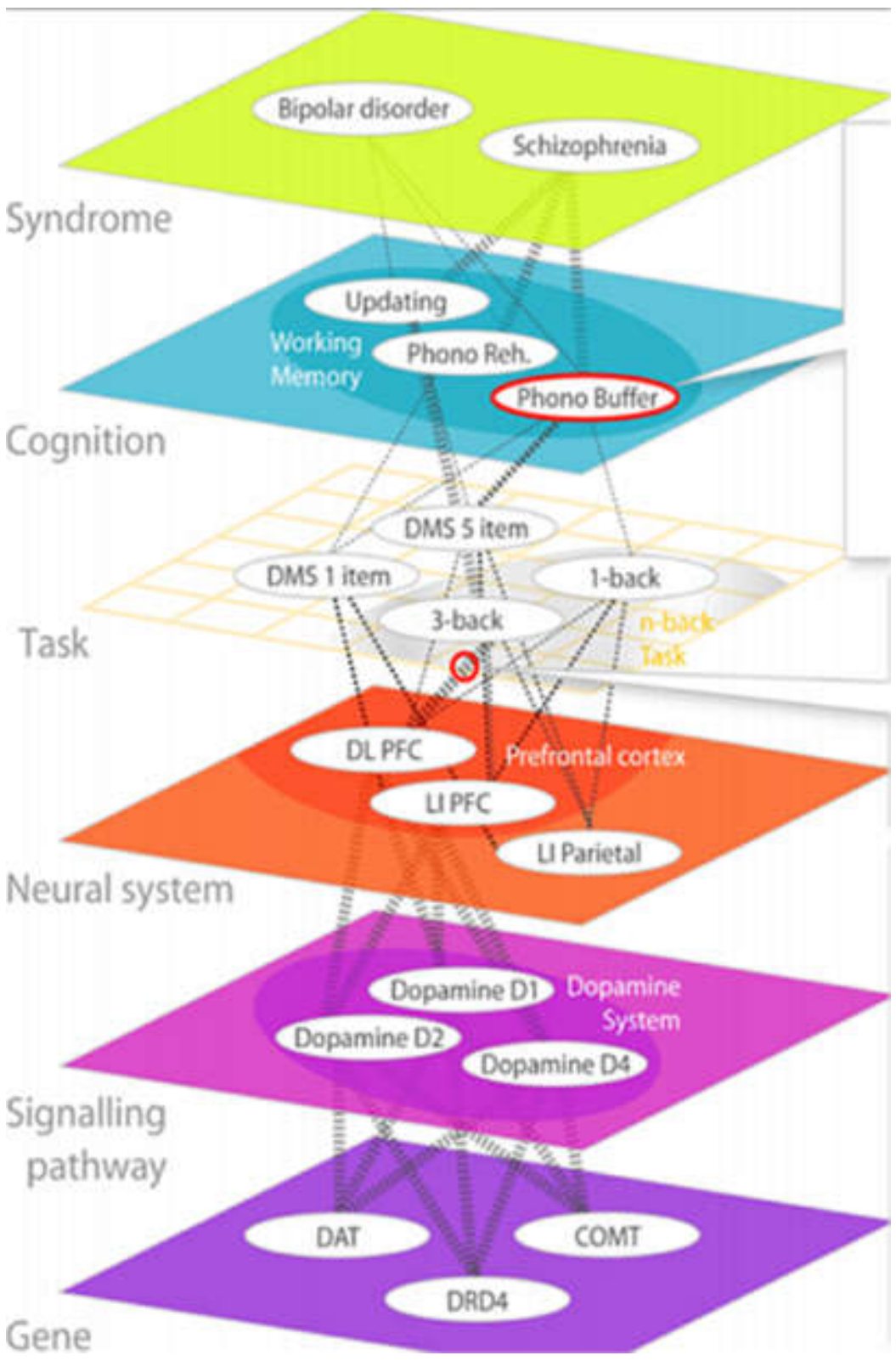

Environment => Genes => Proteins => signaling pathways, receptors, ion channels, synapses => properties of neurons => development of neural networks, connectomes => neurodynamics => cognitive phenotypes => behavior, abnormal behavior => syndromes, mental disorders => interactions with the environment.

I will present here analysis of neurodynamical processes based on computational simulations and fMRI neuroimaging experiments, referring to Lewin's ideas. Artificial neural networks are constructed from computational units representing neurons. They receive signals, do simple calculations and send signals to other neurons. The state of the network is characterized by the pattern of activity of neurons that changes in time. Biological neurons have very complex structures. Neural simulators should take into account at least basic biological properties of neurons, such as excitatory and inhibitory types of connections (activating ion channels in synapses of neurons that let positively and negatively charged ions form currents flowing between inter and extra-cellular space), and spontaneous depolarization decreasing activity of neurons (leak current channels). Neural simulator called "Emergent" is a suitable tool provid-ing biologically inspired model of neurons and their networks (Aisa, Mingus, O'Reilly, 2008). The 3-layer model of reading has separate layers O for orthography, P for phonology, and a large layer S to represent semantics as a distributed activity over 140 neural units representing microfeatures defining concepts. Additional hidden layers transform signals flowing in both directions between layers O ⇔ P, O ⇔ S, and P ⇔ S. The system has learned to map each of the 3 layers to the other two for a set of 40 words by adjusting strength of synaptic connections.

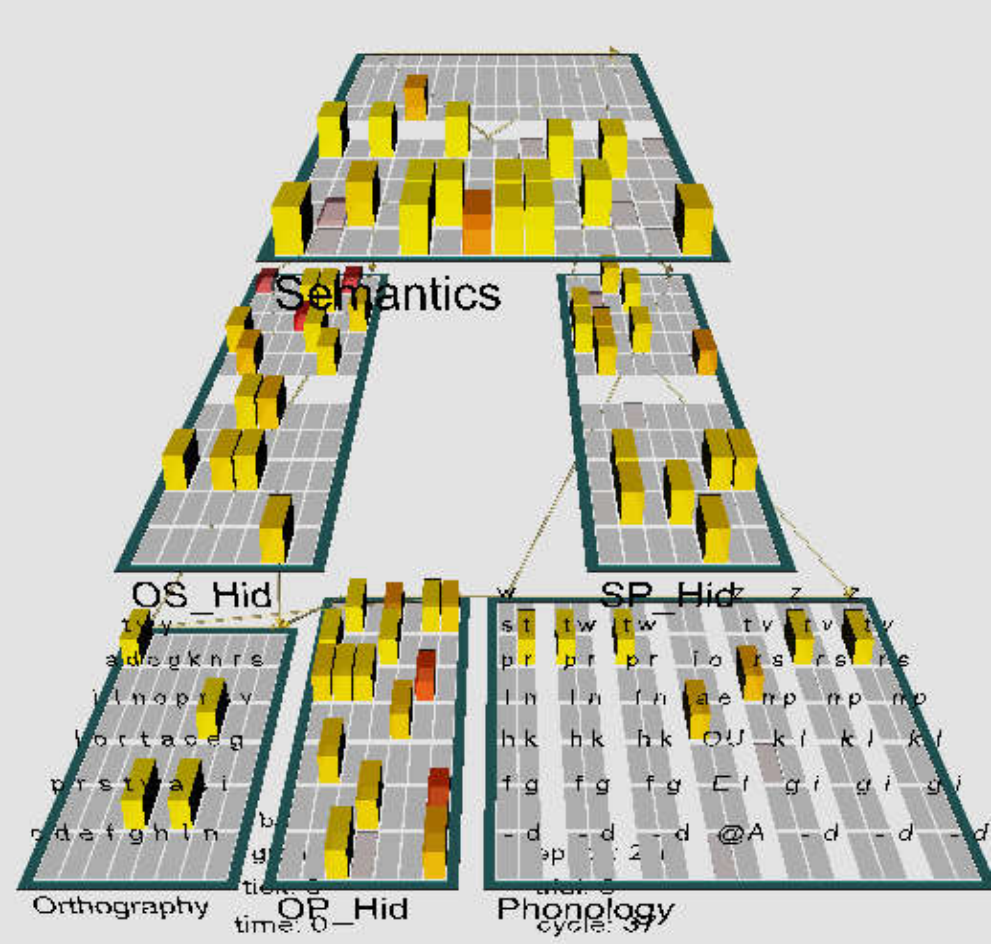

Activation of the orthographic layer will lead to specific patterns of activation in the phonological and semantic layers, etc. These activations are not static, they fluctuate around specific pattern that persists for some time, depending on network properties and noise in the system. Since these quasi-stable patterns attract activity from similar patterns, competing with each other, they are called attractors of the dynamics. All states that are attracted to the same pattern belong to the basin of attraction for that specific pattern. Basins of attractors divide the space of neural activity into distinct regions. In case of semantic layer such patterns represent concepts, and are linked to symbolic representation in O and P layers. Neural networks that have this type of dynamics are called attractor networks.

Transitions between attractor states are possible because neurons active in such states after some time will decrease their activity and desynchronize (due to the leak currents leading to spontaneous depolarization). Also noise in the system and new stimuli may push the system out of the attractor basins. This process has been described in Lewin theory in terms of psychological forces that act on life energy field changing its state. In the model here neurodynamics means changes in 140-dimensional patterns of semantic layer activity. This can be visualized using several techniques: recurrence plots, fuzzy symbolic dynamics or MDS visualization (Duch and Dobosz, 2011). Presenting selected word as input in O or P layer the system reaches attractor state representing the semantics of this word, and the S layer pattern fluctuates staying in the basin of attraction, and after some time making rapid transition to another attractor state. These transitions usually take place between concepts that have weak overlap, sharing some microfeatures. Transitions at the semantic level result in activation of the phonological layer that produce a stream of words, serving as a model of the stream of thoughts. Dwell time in attractor basin determines the speed of changes in the mental field, or speed of attention shifts.

In Fig. 2 each point represents specific activity pattern in 140 dimensions. Recurrence plots (top left) show using color codes distance $D(x(t_i),x(t_j))$ between the current state at time $t_i$ and the state at time $t_j$. Dark square areas along the diagonal show that the system stays in the attractor basin, the trajectory $x(t_i)$ changes only slightly. After a short time there is a fast change to another attractor. Intermediate states during transitions are too short to activate symbolic representations and have no semantic interpretation. Wandering between different attractor states is clearly seen in the MDS representation (top right). Fluctuations of the patterns in the FSD representation show a few points during transition between basins of attractors and dense cloud of points inside the basin of attraction.

This type of analysis shows the speed of attention shifts in semantic layer in reaction to external stimuli or intrinsic dynamics. Certain dysfunctions at the single neuron or neural network level may lead to problems with attentions shifts, they may either be to slow or too fast (hyperactivity), as it is observed in case of autism or ADHD (Duch et al, 2012).

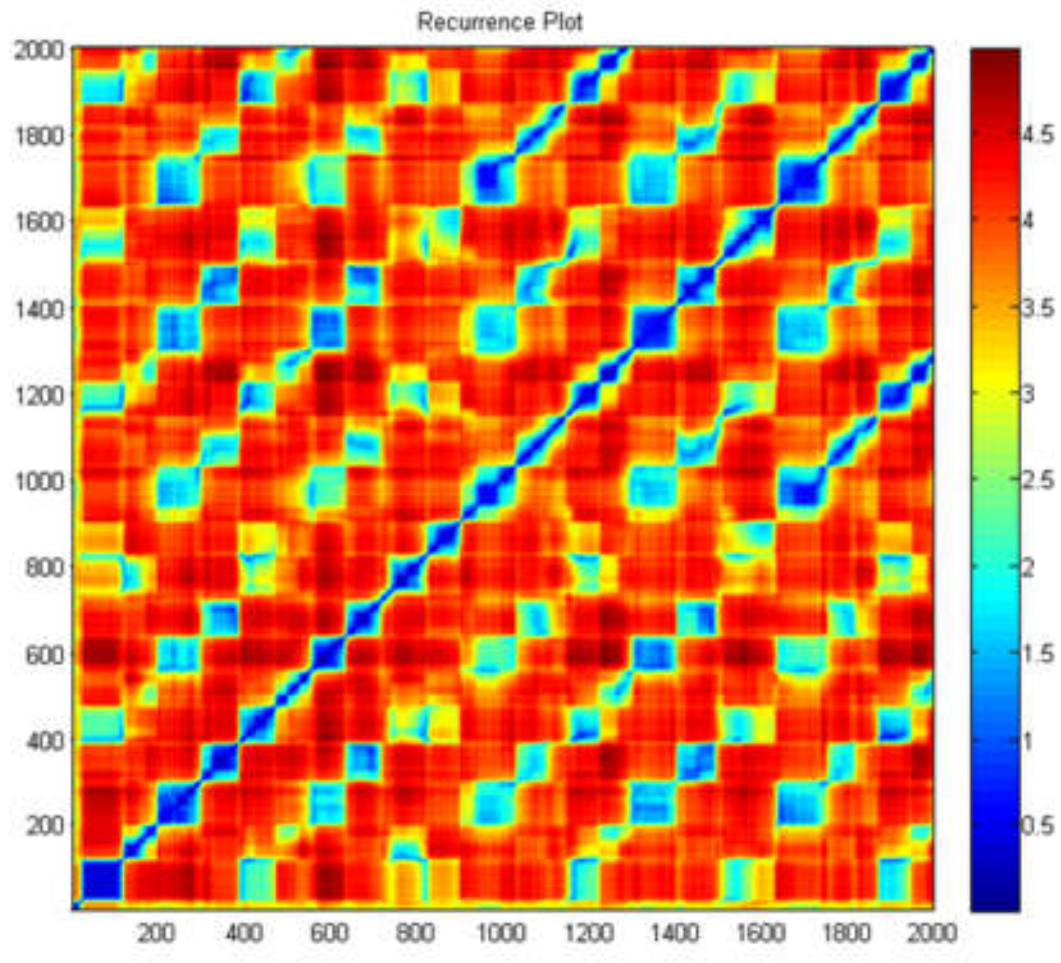

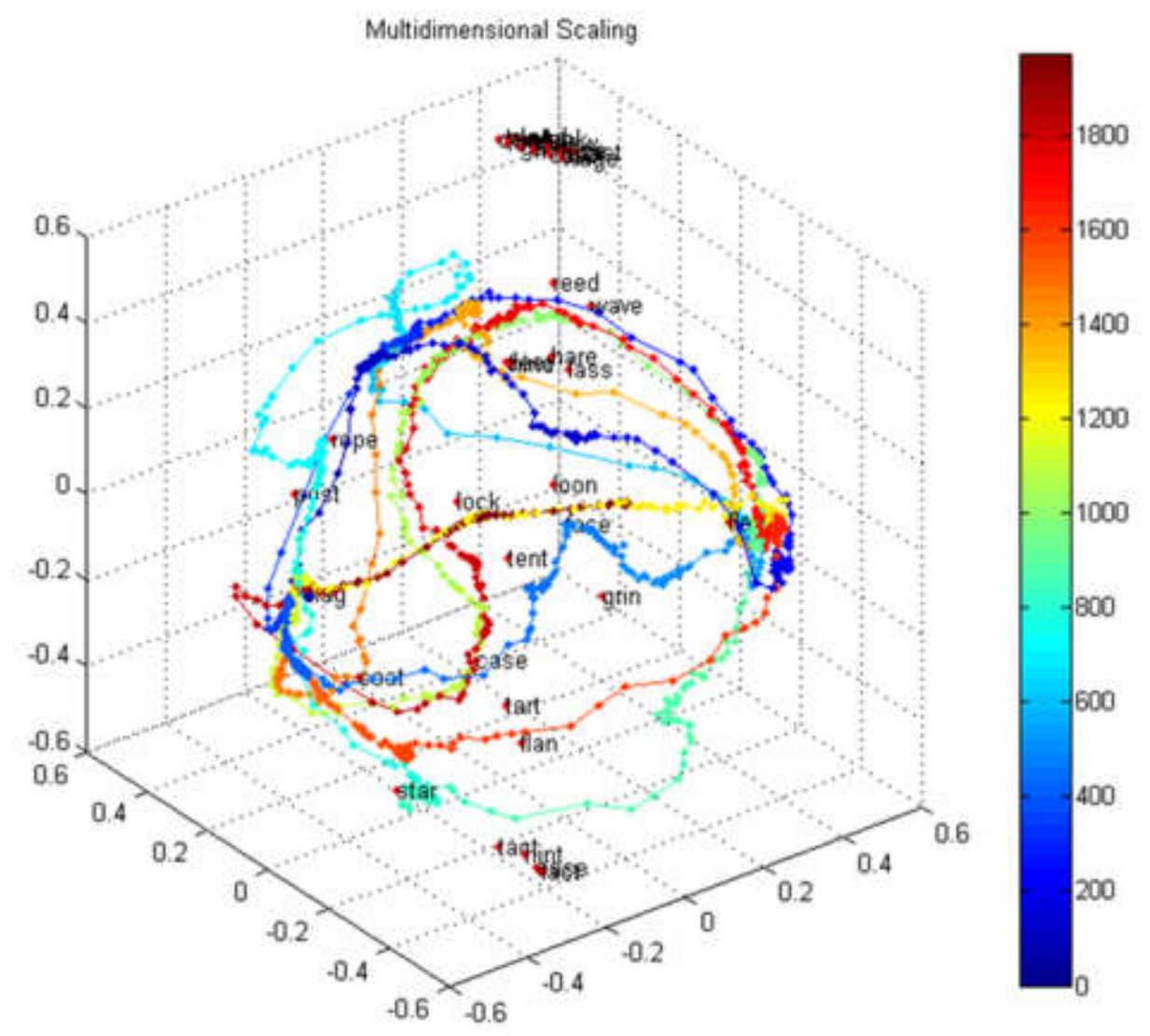

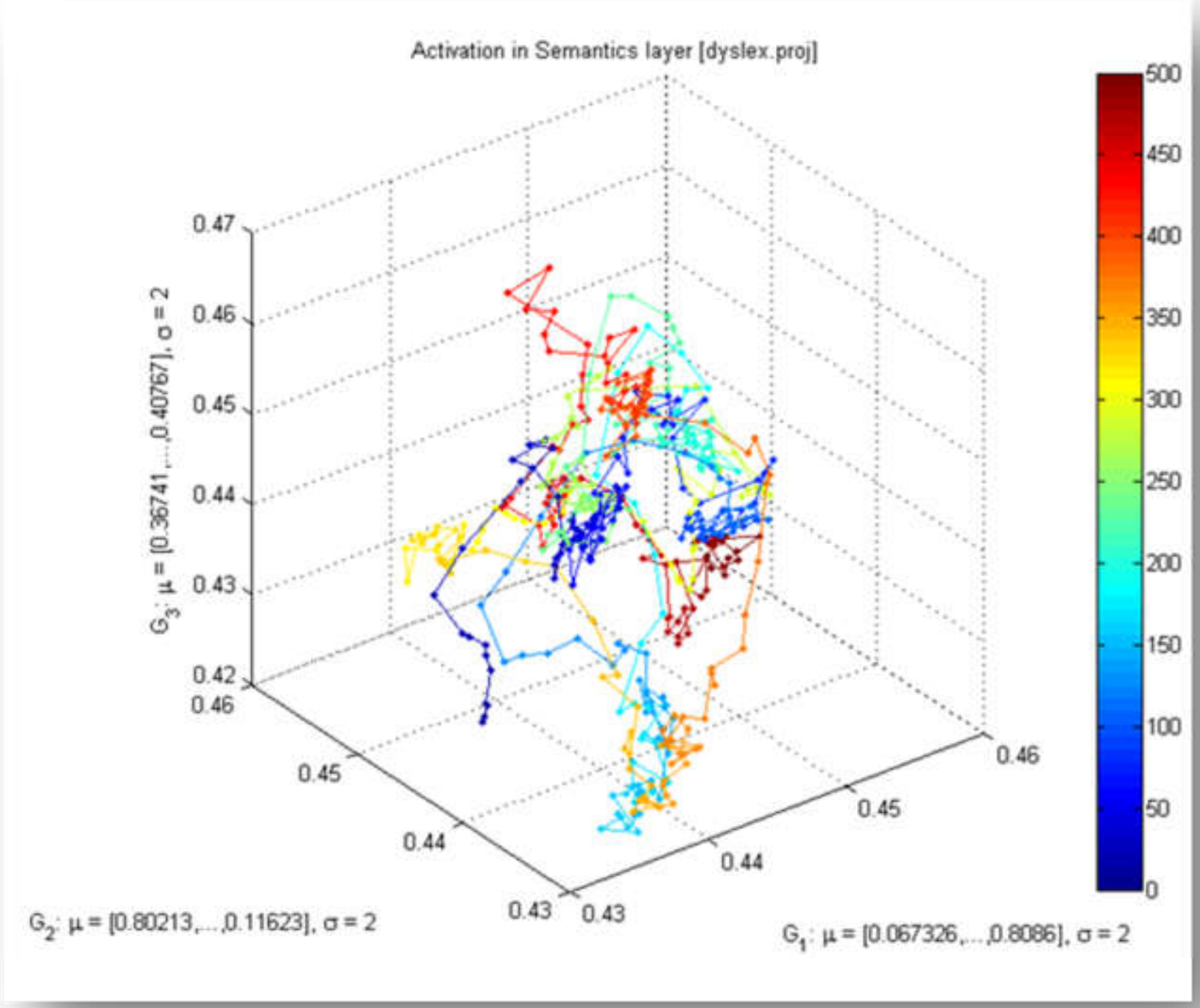

Similar visualization of dynamics is possible with brain signals from EEG, MEG or fMRI. In this case some measures of activity of small brain patches that estimate brain cognitive activity are evaluated. In case of EEG or MEG it can be areas that have large clustering coefficients of Phase Locking Values (PLV) or other measures. In case of fMRI it can be clusters of voxels with increased or decreased activity. In this way patterns of correlated activity are discovered, and knowing brain anatomy they can sometimes be interpreted in a meaningful way. One of the most exciting areas of brain research is network science (Bassett and Sporns, 2017), analysis of active networks in the brain. Networks have nodes (localized populations of neurons) that are activated when specific functions are performed. The best hope for understanding the sources of cognitive activity, elucidating details of cognitive processes, is based on network science. This approach is still in the early stages, our ability to extract meaningful information from brain signals is limited, but it already allowed for asking specific questions that could not be formulated at the behavioral level.

Basic human brain anatomy is similar, with standard divisions between brain lobes, fissures, gyri, and sulci of the cerebral cortex, and subcortical nuclei, but individual variability is high. Are the functional networks similar in all brains, or are they highly individual? Are the brain regions highly specialized or are they flexible, and similar level of competence can be reached using different sets of network nodes? Can the whole-brain network properties change during active task performance? All these questions have important implications not only for understanding cognitive processes, but also for practical applications, for example in neurorehabilitation. The division between automatic and deliberate psychological processes is now commonly accepted. Global Neuronal Workspace Theory (Deahene et al. 1998) assumes two main computational spaces: a set of specialized and modular perceptual, motor, memory, evaluative, and attentional processors, and a unique global workspace composed of distributed and heavily interconnected nodes connected by long-range axons. Thinking, deliberation, problem solving, or in general intelligent behavior in new situations require flexibility at the global workspace level. If the need arises, for example cognitive load on the whole-brain network is high, they may recruit additional brain regions, including regions that are usually active in resting state (Finc et al. 2017). Higher network modularity is correlated with higher working memory capacity and better performance. Strong connectivity within modules and sparse connections between modules increases effective cooperation of brain regions, and is associated with higher IQ. Individual connectome and functional networks that can be activated in this neural space probably determine all personality traits, preferences, and cognitive abilities. They may be used to identify various mental disorders. For example, estimating the strength of the most important 16 functional connections was sufficient to reach 85% accuracy in distinguishing autistic people form the healthy ones (Yahata et al. 2016).

## 4. Conclusions

Psychiatry has been based on constructs that were derived from behavioral syndromes. The attempt to define Research Domain Criteria (RDoC) based on multi-level phenomics shows that traditional approach has exhausted its ability for description of abnormal behavior, because "… these categories, based upon presenting signs and symptoms, may not capture fundamental underlying mechanisms of dysfunction" (Insel et al. 2010). Neurocognitive phenomics based on analysis of major brain networks and dynamical concepts is our best chance for understanding abnormal but also normal behavior. Following psychiatry, it should be actively pursued in psychology and learning sciences (Duch, 2013). Such approach will confirm the basic insights of Kurt Lewin, linking psychology with neuroscience. Without relating psychological constructs to brain processes situation will be similar to parametric theories that explain sunsets and sunrises by fitting models to the data. Psychological constructs should "capture fundamental underlying mechanisms", and that requires understanding of neurodynamics.

Does it mean that classical psychological concepts should be eliminated, as Paul Churchland (1981) claimed? This is highly doubtful. Approximate description of causal structure of brain or behavioral states may sometimes be more informative at a macroscale, as shown by information theory analysis (Hoel, 2017). Complex psychological concepts certainly need to be better aligned with neurocognitive phenomics. That includes self, personality, consciousness, intelligence, talent. Even in categorization experiments purely psychological explanations may be quite different than those based on nonlinear dynamics of neural networks (Duch, 1996). Neurodynamics and neurocognitive phenomics are the key for further development of psychological constructs. Is there a shorter route to deep understanding of human behavior?

**Acknowledgements**: This work has been supported by the National Science Center grant UMO-2016/20/W/NZ4/00354.